\documentclass[12pt,a4paper,dvips]{article}
\usepackage{a4p}
\usepackage{cite,mcite}
\usepackage{graphicx}
\usepackage{physics}
\usepackage{l3_title,ifthen}
\journalname{Phys. Lett. B} 

\date{February 6, 2002}
\preprint{2002-010}

\newcommand{\LL}{\ell^+ \ell^-}
\newcommand{\EE}{\rm e^+ e^-}
\newcommand{\MM}{\mu^+ \mu^-}
\newcommand{\TT}{\tau^+\tau^-}
\newcommand{\QQ}{\rm q \bar{q}}

\newcommand{\WW}{\rm W^+ W^-}

\newlength{\capindent}
\newlength{\capwidth}
\newlength{\figwidth}
\newcommand{\icaption}[2][!*!,!]{\hspace*{\capindent}%
  \begin{minipage}{\capwidth}
    \ifthenelse{\equal{#1}{!*!,!}}%
      {\caption{#2}}%
      {\caption[#1]{#2}}
  \end{minipage}}
\begin{document}
\bibliographystyle{l3style}
\begin{titlepage}
\title{ 
Search for a Higgs Boson\\ Decaying into Two Photons at LEP}
\author{The L3 Collaboration}
%
%

\begin{abstract}
A Higgs particle produced in association with a Z boson and decaying into 
two photons
is searched for in 
the data collected by the L3 experiment at LEP.
All possible decay modes of the Z boson are investigated. No signal is 
observed in
447.5 pb$^{-1}$ of data recorded at centre-of-mass energies from 192$\GeV$ up to 209$\GeV.$  
Limits on the branching fraction of the Higgs boson decay
into two
photons as a function of the Higgs mass are derived. A lower limit on the 
mass of a
fermiophobic Higgs boson is set at 105.4$\GeV$ at 95$\%$ confidence level.
\end{abstract}
%
%
\submitted
\vspace*{20mm}
\end{titlepage}
\section*{Introduction}

The Standard Model of the electroweak interactions allows the decay of a Higgs boson h 
into a photon pair 
only at the one loop level.  
The branching fraction of this decay is small \cite{sm}. It amounts to about 0.1$\%$ for  
Higgs masses $m_{\mbox{{\scriptsize h}}}$ between 80 and 110$\GeV$.
Several extensions of the Standard Model
predict enhancements of this branching fraction \cite{om}. 
For instance, an appropriate choice of the parameters of the 
Two Higgs Doublet Models of Type~I~\cite{thdm}, predicts
the lightest CP even Higgs not to couple to fermions at tree level.  
Such a Higgs is expected to decay dominantly into a pair of
photons if its mass is below 90$\GeV$ \cite{fphob}.
In Higgs triplet models, one of the neutral scalars may have a large branching fraction
BR(h $\to \gamma \gamma$) and be produced at LEP with rates comparable
to the Standard Model ones \cite{triplet}. At higher masses, the
branching fraction BR(h $\to \gamma \gamma$) decreases to the
advantage of the W$^\pm$W$^{\mp *}$ mode.

\indent
This letter presents the search for a Higgs boson produced in association with a Z boson 
through the process $\mbox{e}^{+}\mbox{e}^{-}\to$ Zh, followed by the decay h $\to \gamma \gamma$.
All decay modes of the \mbox{Z} boson are investigated. 

\indent
The high energy data sample collected by the L3 detector \cite{l3det} at centre-of-mass $\sqrt s$
up to 209$\GeV$ is analysed.
Results from partial samples were previously reported by L3\cite{oldl3,l3}
and other LEP Collaborations \cite{aleph}.

\section*{Data and Monte Carlo Samples}

We analyse
data collected with the L3 detector 
during the years 1999 and 2000 
at centre-of-mass energies $\sqrt s=192-209\GeV$, for a total 
integrated luminosity of 447.5 pb$^{-1}$. The data are grouped into ten samples whose average centre-of-mass energies
and corresponding integrated luminosities are listed in Table \ref{tab_lumi}.
Results from 176 pb$^{-1}$ of integrated luminosity collected 
at $\sqrt s$ = 189$\GeV$ are included in the final results.

The Standard Model Higgs production cross section is calculated 
using the HZHA generator~\cite{hzha}. 
Signal Monte Carlo samples are 
generated using PYTHIA \cite{pythia} for Higgs masses between 50 
and 120$\GeV$. These samples comprise between 500 and 2000 signal events 
depending on the search channel. 
For background studies the following Monte Carlo programs are used:
KK2f \cite{kk2f} ($\EE \to \QQ (\gamma)$, $\EE \to \MM (\gamma)$, $\EE \to \TT (\gamma)$), 
PYTHIA ($\EE \to \rm ZZ$ and $\EE \to \rm Z\EE$),
KORALW \cite{koralw} ($\EE \to \WW$), 
PHOJET \cite{phojet} ($\EE \to \EE \QQ$),
KORALZ \cite{koralz} ($\mbox{e}^{+}\mbox{e}^{-} \to \nu \bar{\nu} (\gamma)$),
GGG \cite{ggg} ($\mbox{e}^{+}\mbox{e}^{-} \to \gamma \gamma (\gamma)$),
BHWIDE \cite{bhwide} ($\mbox{e}^{+}\mbox{e}^{-} \to \mbox{e}^{+}\mbox{e}^{-} (\gamma)$),
TEEGG \cite{teegg} ($\mbox{e}^{+}\mbox{e}^{-} \to \mbox{e}^{+}\mbox{e}^{-} \gamma$), 
DIAG36 \cite{diag36} ($\mbox{e}^{+}\mbox{e}^{-} \to 
\mbox{e}^{+}\mbox{e}^{-}\mbox{e}^{+}\mbox{e}^{-}$) and
EXCALIBUR \cite{excalibur} for other four-fermion final states. 
The number of simulated events for the most 
important background channels is at least 100 times higher than the corresponding 
number of expected events in the data. 

The L3 detector response is simulated using the GEANT \cite{geant}
program, which takes into account the effects of energy loss,
multiple scattering and showering in the detector. The GHEISHA \cite{gheisha} package
is used to simulate hadronic interactions in the detector.
Time dependent inefficiencies, as monitored during the data acquisition period, are also
simulated.

\section*{Analysis Procedure}

The analysis aims to select events with isolated photons
and a Z boson.
Three final states are then investigated, according to the Z boson decay. They are denoted as
$\mbox{q} \bar{\mbox{q}} \gamma \gamma $,
$\nu \bar{\nu} \gamma \gamma$ and $\LL \gamma \gamma$, with $\ell =\mbox{e},\mu,\tau$.
The selections for each final state are described
in the following and proceed from a common photon identification.

Photons are selected from clusters in the BGO electromagnetic calorimeter
with an energy greater than 1$\GeV$ and a shower profile compatible
with that expected for an electromagnetic particle. The ratio of the energies
measured in a 3$\times$3   
and a 5$\times$5 matrix, centred on the most energetic crystal,
must exceed 0.95.
The energy deposit in the hadron calorimeter is required to be below 20$\%$ of
the energy measured in the electromagnetic calorimeter.

\begin{figure}[H]
\begin{center}
\begin{tabular}{l}
\includegraphics[width=17cm]{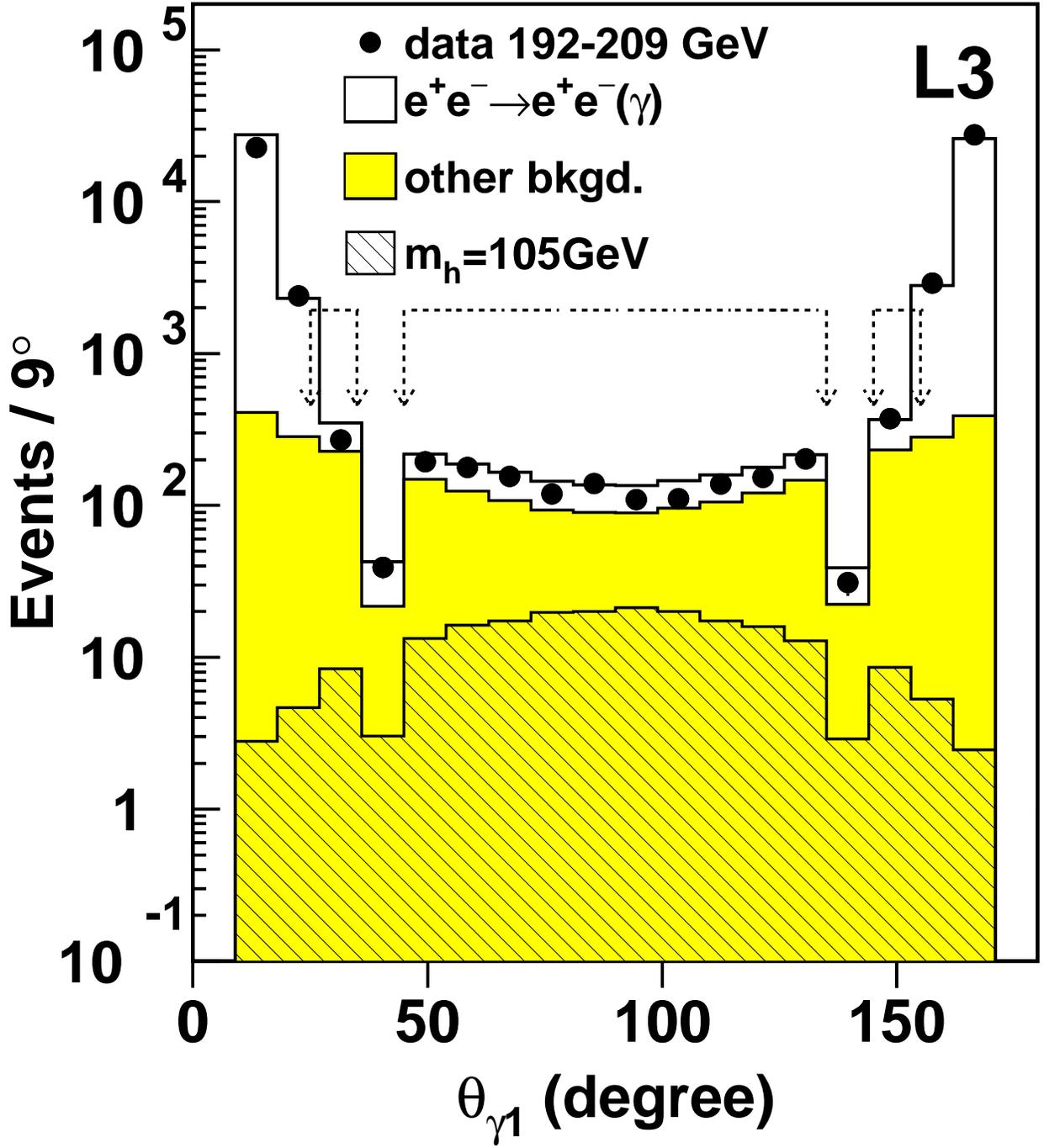} \\  
\end{tabular}
\caption{Distribution of  
the polar angle, $\theta_{\gamma_1}$ of the most energetic photon
for  data and background. A
Higgs boson signal with mass $m_{\mbox{{\scriptsize h}}}$ =  105$\GeV$ is superimposed with  arbitrary normalisation.
All Z final states are combined. The selected regions are indicated by the arrows.
}
\label{fig_polar}
\end{center}
\end{figure}

No track should point to the clusters within 50 mrad
in the plane perpendicular to the beam axis. 
Only clusters in the polar angle ranges $25^{\circ}< \theta <35^{\circ}$, 
$45^{\circ}< \theta <135^{\circ}$ and $145^{\circ}< \theta <155^{\circ}$
are considered, well within the coverage of
the barrel and end-cap regions of the BGO electromagnetic calorimeter. 
The distribution of the polar angle  
of the most energetic photon is shown in Figure \ref{fig_polar}.

These analyses require at least two photons.
Photons from the decay of a heavy resonance are expected to be relatively energetic, hence
the energy of the hardest photon has to exceed 
10$\GeV$  and the energy of the second most energetic photon has to exceed 6$\GeV.$

\subsection*{The $\mbox{q} \bar{\mbox{q}}${\boldmath $\gamma \gamma$} Final State}

\begin{figure}[H]
\begin{center}
\begin{tabular}{l}
\includegraphics[width=17cm]{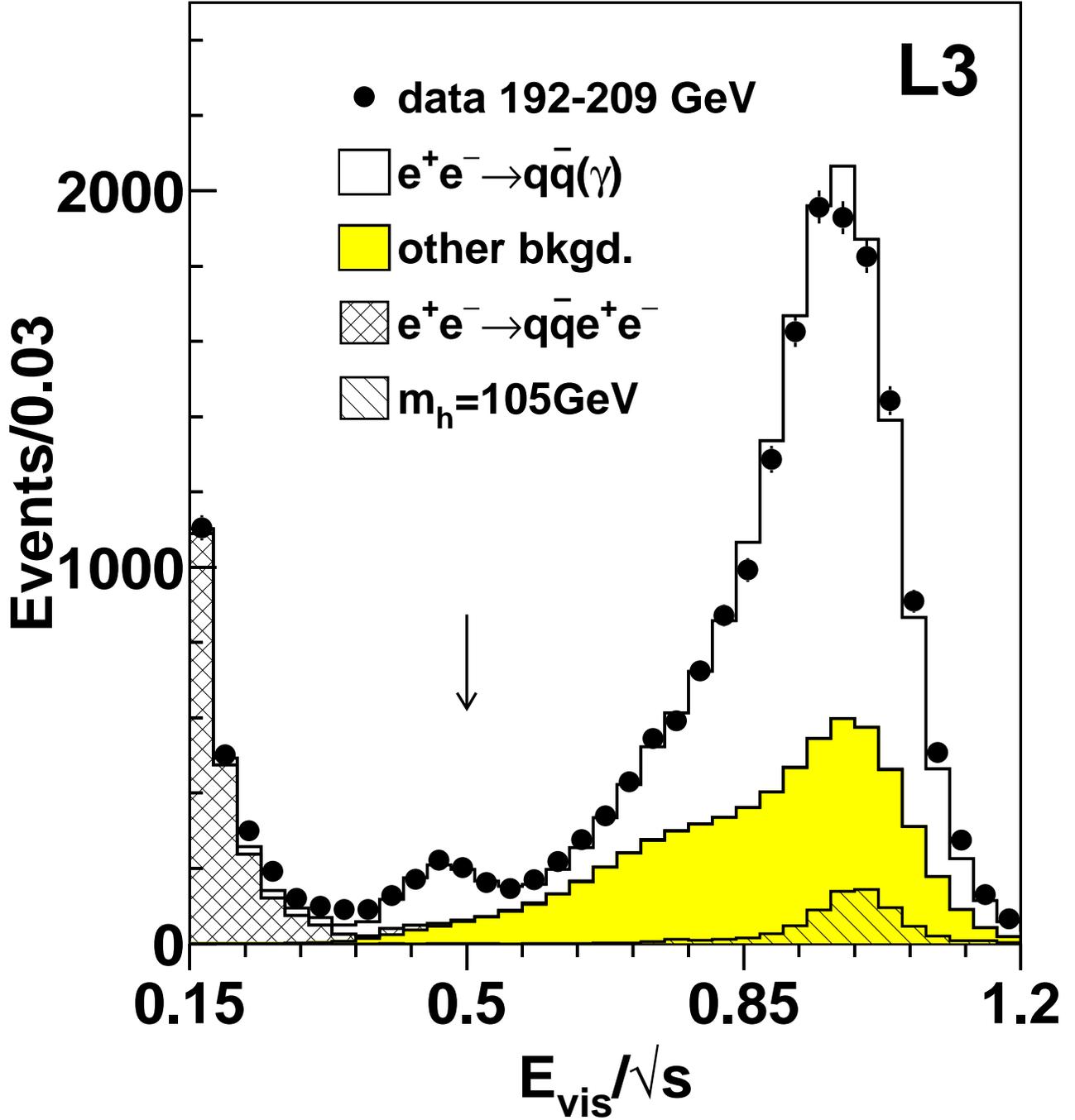} \\  
\end{tabular}
\caption{Distribution of $E_{vis}/\sqrt{s}$ after the hadronic preselection
for data and background for the $\mbox{q} \bar{\mbox{q}} \gamma \gamma$ final state. A
Higgs boson signal with mass $m_{\mbox{{\scriptsize h}}}$ =  105$\GeV$
is superimposed with  arbitrary normalisation.
The arrow indicates the value of the cut.
}
\label{fig_visib}
\end{center}
\end{figure}

Candidates in the $\mbox{q} \bar{\mbox{q}} \gamma \gamma$ final state are characterised by a
pair of isolated photons accompanied by two jets. The 
$\mbox{q} \bar{\mbox{q}} \gamma \gamma$
selection starts from a preselection of hadronic events: 
high multiplicity events are required to be balanced and their visible energy $E_{vis}/\sqrt{s}$
has to be larger than 0.5.
The background 
from two-photon interaction events is reduced requiring the energy in a 30$^{\circ}$
cone around the beam pipe to be less than half of the visible energy.
The preselection yield is reported in Table \ref{tab_sel1}. Figure \ref{fig_visib}
shows the comparison between data and the Monte Carlo expectations for the distribution of
$E_{vis}/\sqrt s$ for the preselected events.

Events which contain at least two photons  are selected.
In addition,
photons
coming from neutral hadron decays are rejected by requiring the energy in a
10$^{\circ}$ cone around the photon direction to be less than 2.5$\GeV$, and that in 
a 20$^{\circ}$ cone to be less than 4.5$\GeV$. The number of charged tracks and calorimeter 
clusters in a 20$^{\circ}$ cone around the photon direction must be below four.
The opening angle between the photons must be larger than 50$^{\circ}$. 
All other particles are clustered into two jets using
the DURHAM jet algorithm \cite{durham}
and no jet is allowed
within 25$^{\circ}$ around any photon.

\begin{figure}[H]
\begin{center}
\begin{tabular}{l}
\includegraphics[width=9.5cm]{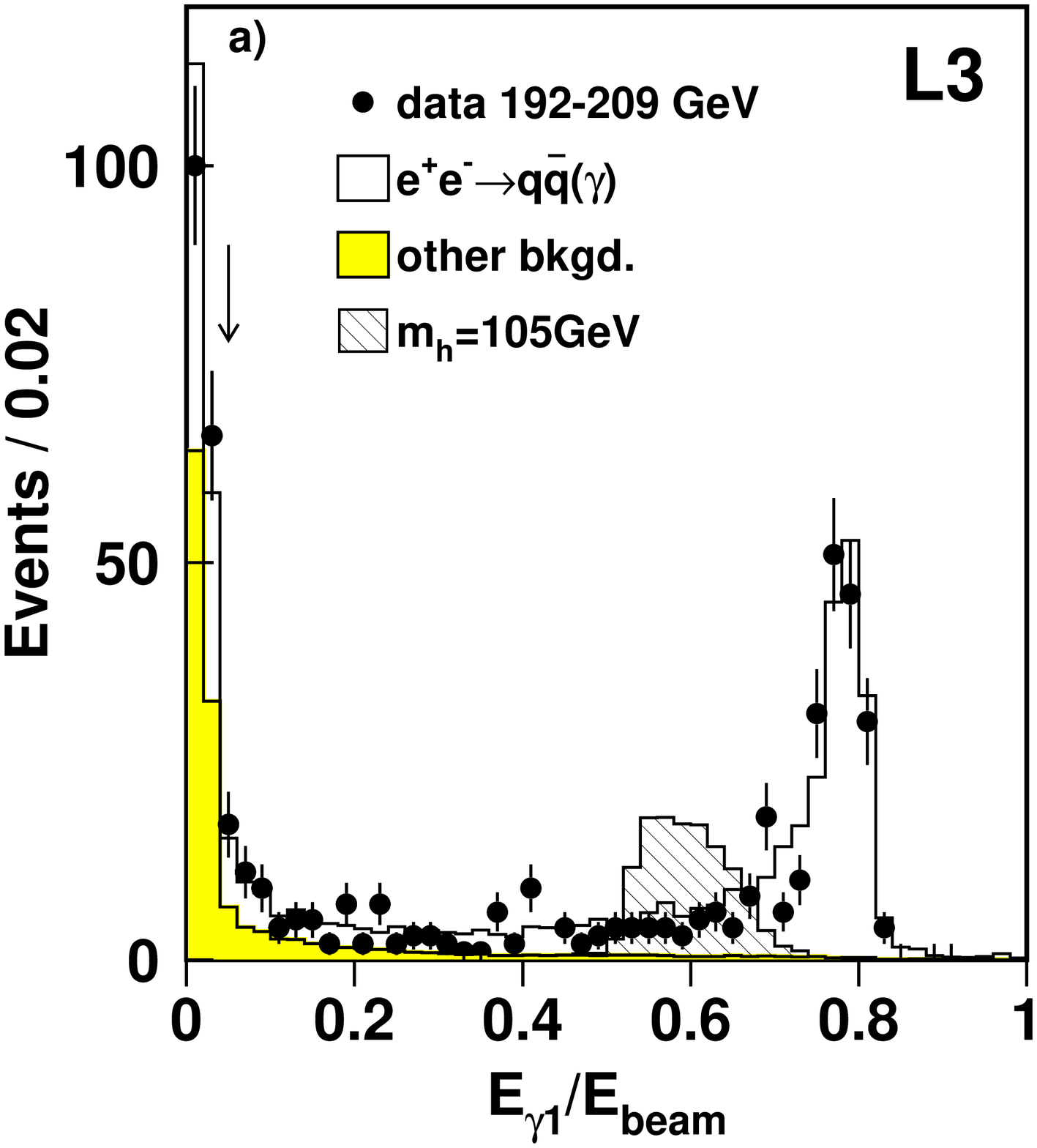} \\
\vspace*{-2cm}
\includegraphics[width=9.5cm]{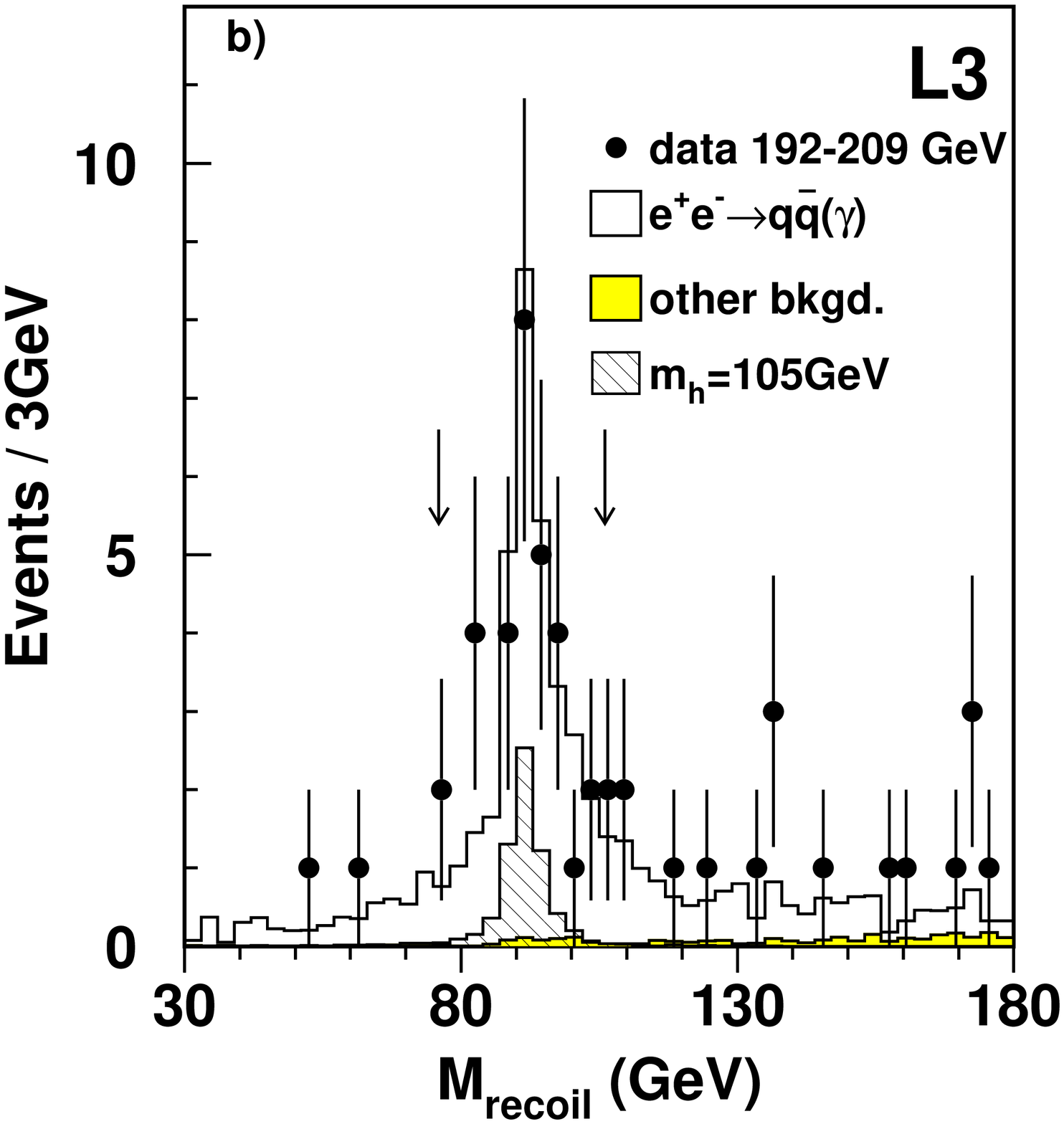} \\
\end{tabular}   
\vspace*{2cm}
\caption{
Distributions for the $\mbox{q} \bar{\mbox{q}} \gamma \gamma$ final state of 
a) the energy of the most energetic photon normalised to the beam energy
and b) the recoil mass against the di-photon system    
in data, background and for a 105$\GeV$
Higgs boson signal with arbitrary normalisation.
The arrows indicate the values of the applied cuts.
}
\label{fig_hadr}
\end{center}
\end{figure}

The energy spectrum of the most energetic photon, normalised to the beam energy, before the application of any 
selection requirement on the 
photon energies, 
is shown in Figure \ref{fig_hadr}a.
Figure \ref{fig_hadr}b presents the distribution of the recoil mass against the di-photon
system after the cuts on the photon energies. 

Finally, the recoil mass against the di-photon system is required to be consistent with the Z mass
within 15$\GeV$.  The efficiency is 40$\%$ for a Higgs boson mass of 100$\GeV$ produced at $\sqrt s$ = 192$\GeV$, 
and 47$\%$ for a Higgs of 110$\GeV$ produced at $\sqrt s$ = 208$\GeV$.
The selection yield  is presented in Table \ref{tab_sel1}. 28 events are observed in the data, with 31 expected from Monte Carlo, mainly from the 
$\mbox{e}^{+}\mbox{e}^{-} \to \mbox{q} \bar{\mbox{q}} (\gamma)$ process.

\begin{figure}[H]
\begin{center}
\begin{tabular}{l}
\includegraphics[width=17cm]{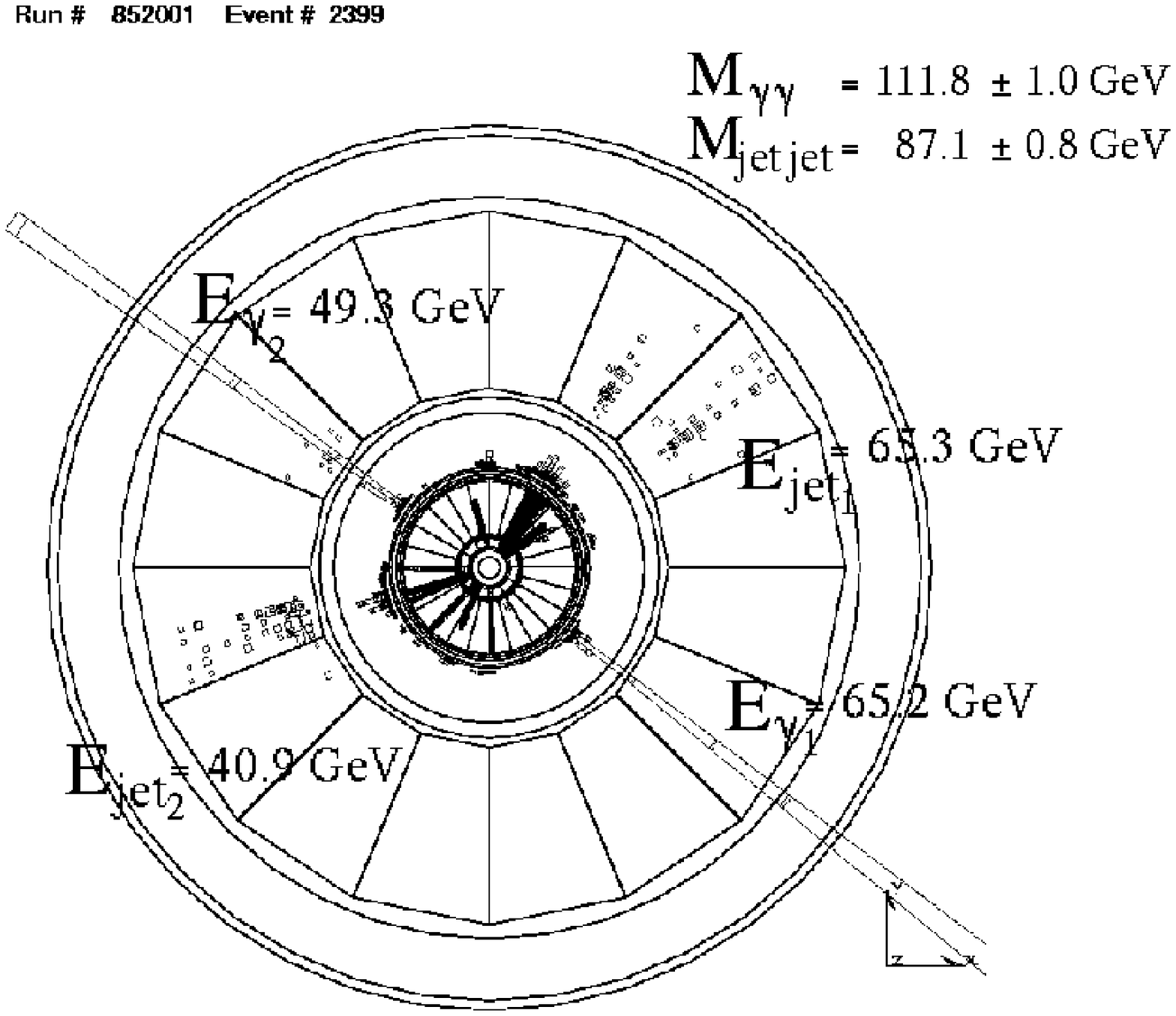} \\  
\end{tabular}
\caption{The   $\EE\to$Zh$\to \mbox{q} \bar{\mbox{q}} \gamma \gamma$  
candidate with the highest di-photon invariant mass.
}
\label{fig_event}
\end{center}
\end{figure}

The event with the highest value of the di-photon invariant mass is displayed in Figure \ref{fig_event}.
It was collected at $\sqrt s$ = 205.1$\GeV$ and its di-photon invariant mass is 111.8 $\pm$ 1.0$\GeV$
while the recoil mass against 
the di-photon system is 87.1 $\pm$ 0.8$\GeV$.

\subsection*{The {\boldmath $\nu \bar{\nu} \gamma \gamma$} Final State}

\begin{figure}[H]
\begin{center}
\begin{tabular}{l}
\includegraphics[width=9.5cm]{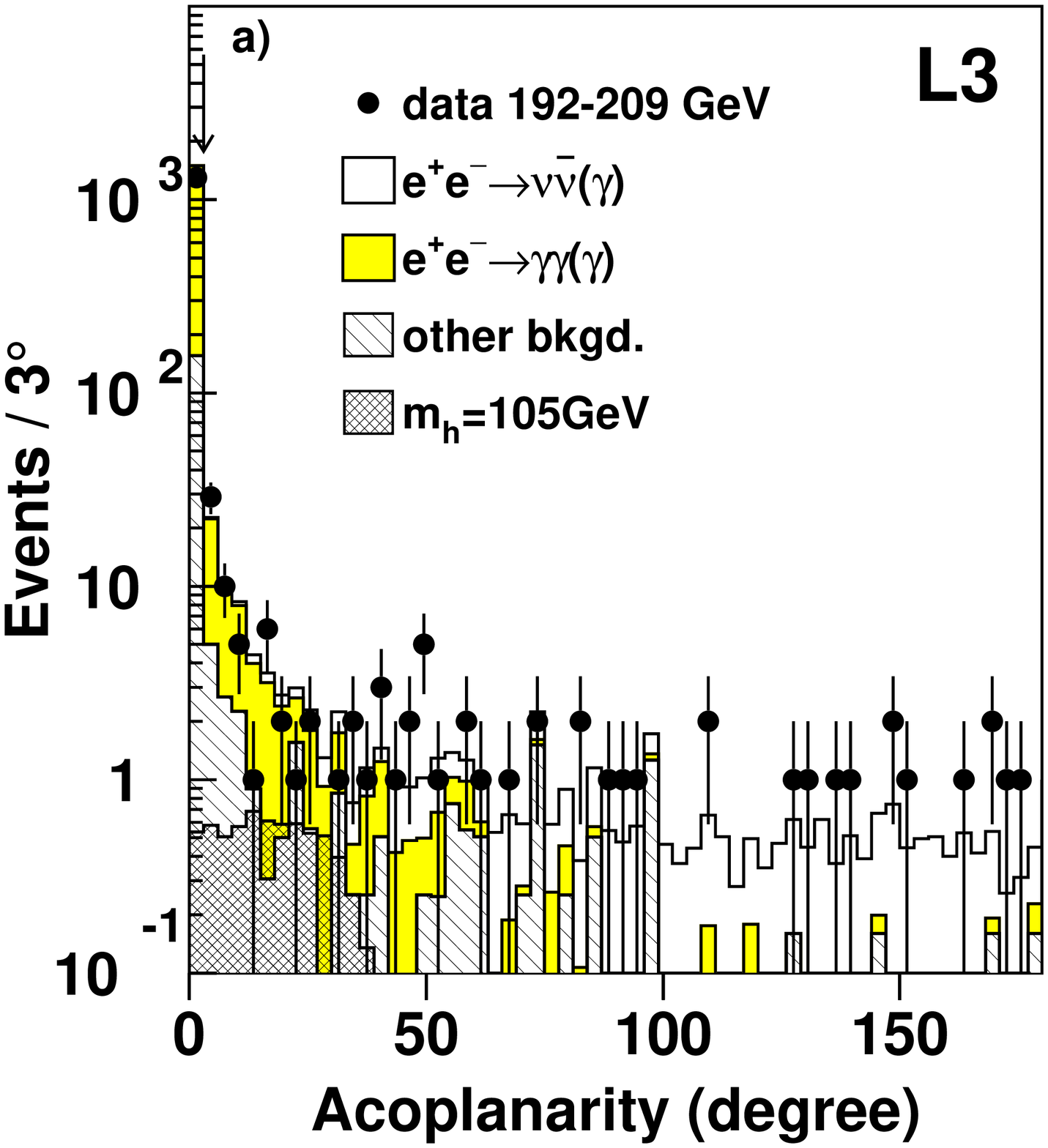} \\
\vspace*{-2cm}
\includegraphics[width=9.5cm]{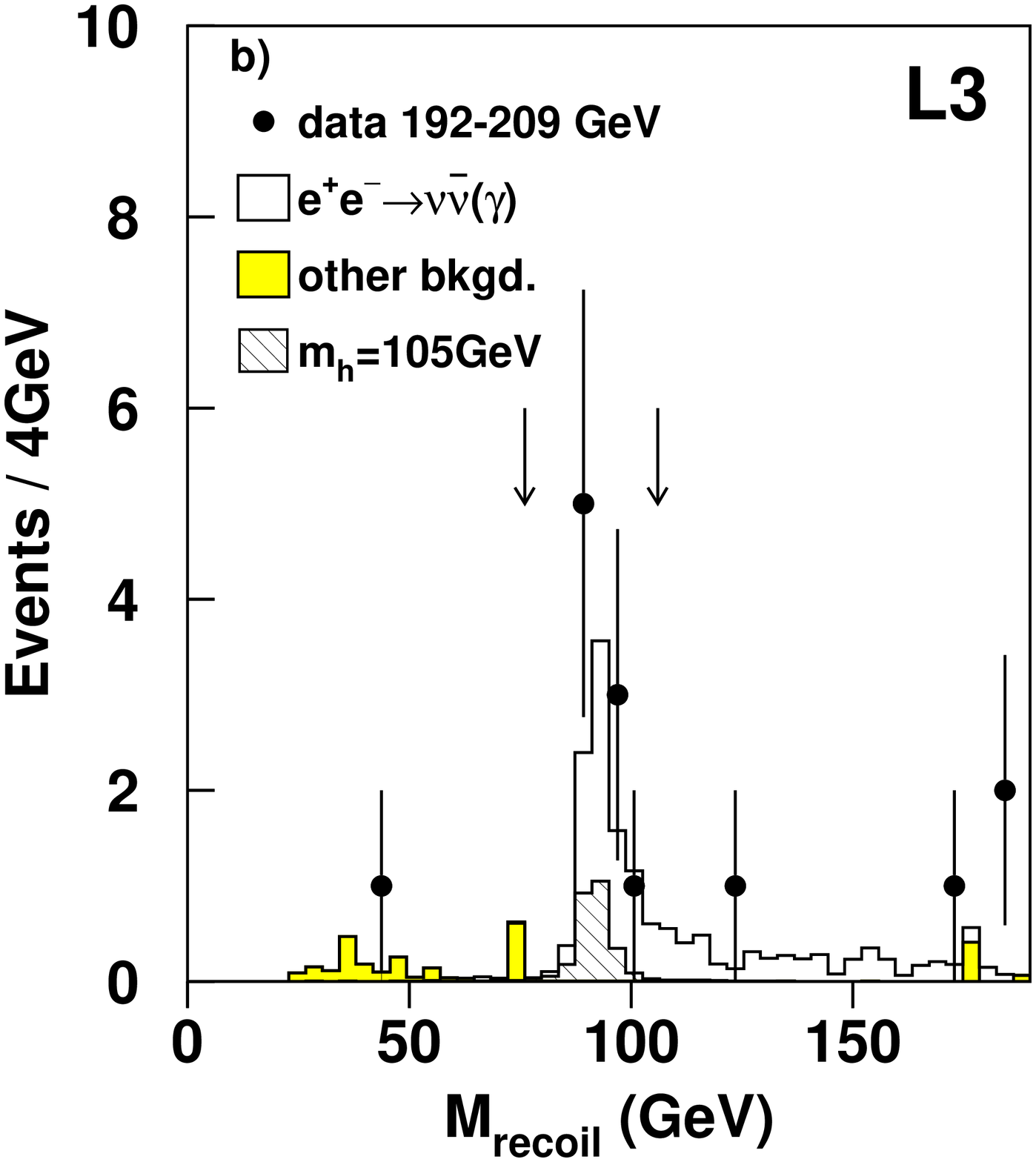} \\
\end{tabular} 
\vspace*{2cm}  
\caption{Distributions for the $\nu \bar{\nu} \gamma \gamma$ final state of 
a) the acoplanarity of the $\gamma \gamma$ system
and b) the recoil mass against the two photons
in data, background and for a
Higgs boson signal with mass $m_{\mbox{{\scriptsize h}}}$ =  105$\GeV$ with arbitrary normalisation.
The arrows indicate the values of the cuts.
}
\label{fig_miss}
\end{center}
\end{figure}

The signature of the 
$\nu \bar{\nu} \gamma \gamma$ final state consists in two 
photons and missing energy. Events are selected that have an identified photon pair, 
no charged tracks and an additional energy below 10$\GeV$. 
To ensure that the missing momentum is well contained in the detector,
the absolute value of the cosine of its
polar angle must not exceed 0.96.
To reduce contributions from the $\mbox{e}^{+}\mbox{e}^{-}\to \gamma \gamma (\gamma)$ process and
from double radiative events with final state particles escaping detection,
the photon acoplanarity is required to exceed 3$^{\circ}$. 
The distribution of this acoplanarity
for the data and the Monte Carlo predictions is presented 
in Figure \ref{fig_miss}a.
The total transverse momentum of the di-photon system must be greater than 2$\GeV$.

Figure \ref{fig_miss}b shows the distribution of the recoil mass against the two most 
energetic photons after  selection requirements on all the other variables.
As final selection criterium,
this mass has to be consistent with the Z boson mass within 15$\GeV.$ The efficiency is 47$\%$ for a Higgs boson mass of 100$\GeV$ produced at $\sqrt s$ = 192$\GeV$ 
and 51$\%$ for a Higgs of 110$\GeV$ produced at $\sqrt s$ = 206$\GeV$.
The number of selected events in data is 9 and 9.2 events are expected from 
the $\mbox{e}^{+}\mbox{e}^{-}\to \nu \bar{\nu} (\gamma)$ process. Other backgrounds are negligible.

\subsection*{The {\boldmath $\LL $}{\boldmath $\gamma \gamma$} Final State}

The $\LL \gamma \gamma$ final state has the characteristic signature of
a photon pair and a lepton pair.
Its selection proceeds from low multiplicity events with two identified photons and an associated 
lepton pair, selected as follows.
 
Electrons are identified from clusters in the electromagnetic calorimeter with an energy greater than
3$\GeV$ and associated to a charged track. The energy deposit in the hadron
calorimeter must be consistent with the tail of an electromagnetic shower. 
Less than 3$\GeV$ are allowed in the electromagnetic calorimeter
in a 10$^{\circ}$ cone around the electron direction. To increase efficiency, events
with just one identified electron are also accepted.
 
Muons are identified from tracks in the muon chambers with a distance to the interaction vertex in
the $r-\phi$ plane below 300 mm and a momentum above 3$\GeV$.
The calorimetric energy in a 10$^{\circ}$ cone around the muon direction must not
exceed 3$\GeV$.
Events with one muon and one minimum ionising particle in the calorimeters are also accepted, 
as well as events with a single muon. 
The background from cosmic rays is rejected by requiring at least one hit in the scintillation 
counters in a 5\,ns
window around the beam crossing time.

Taus are identified as jets 
with one or three tracks in
a 10$^{\circ}$ cone with an energy above 3$\GeV$. The energy in the 
10$^{\circ}-
30^{\circ}$ 
cone around the tau direction has to be below
30\% of the energy in the $0^{\circ}-10^{\circ}$ cone. 
Events with only one identified tau lepton are also accepted.

The preselection of $\LL \gamma \gamma$ events yields the results listed  
in Table \ref{tab_sel2}.
Double radiative di-lepton events are the dominant background and are rejected further 
by requiring the energy of the most energetic lepton to be below 80$\GeV$.

\begin{figure}[H]
\begin{center}
\begin{tabular}{l}
\includegraphics[width=9.5cm]{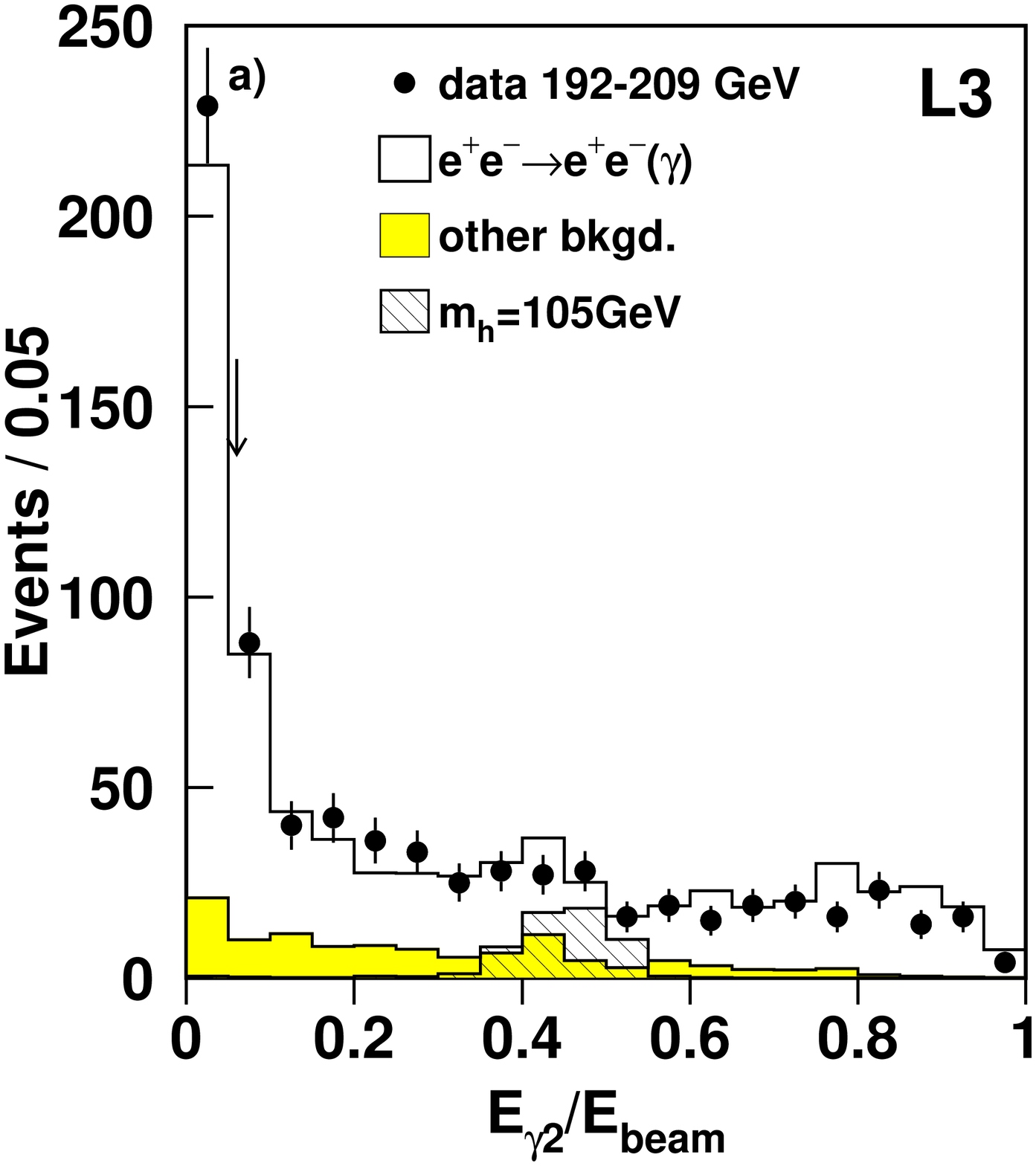} \\
\vspace*{-2cm}
\includegraphics[width=9.5cm]{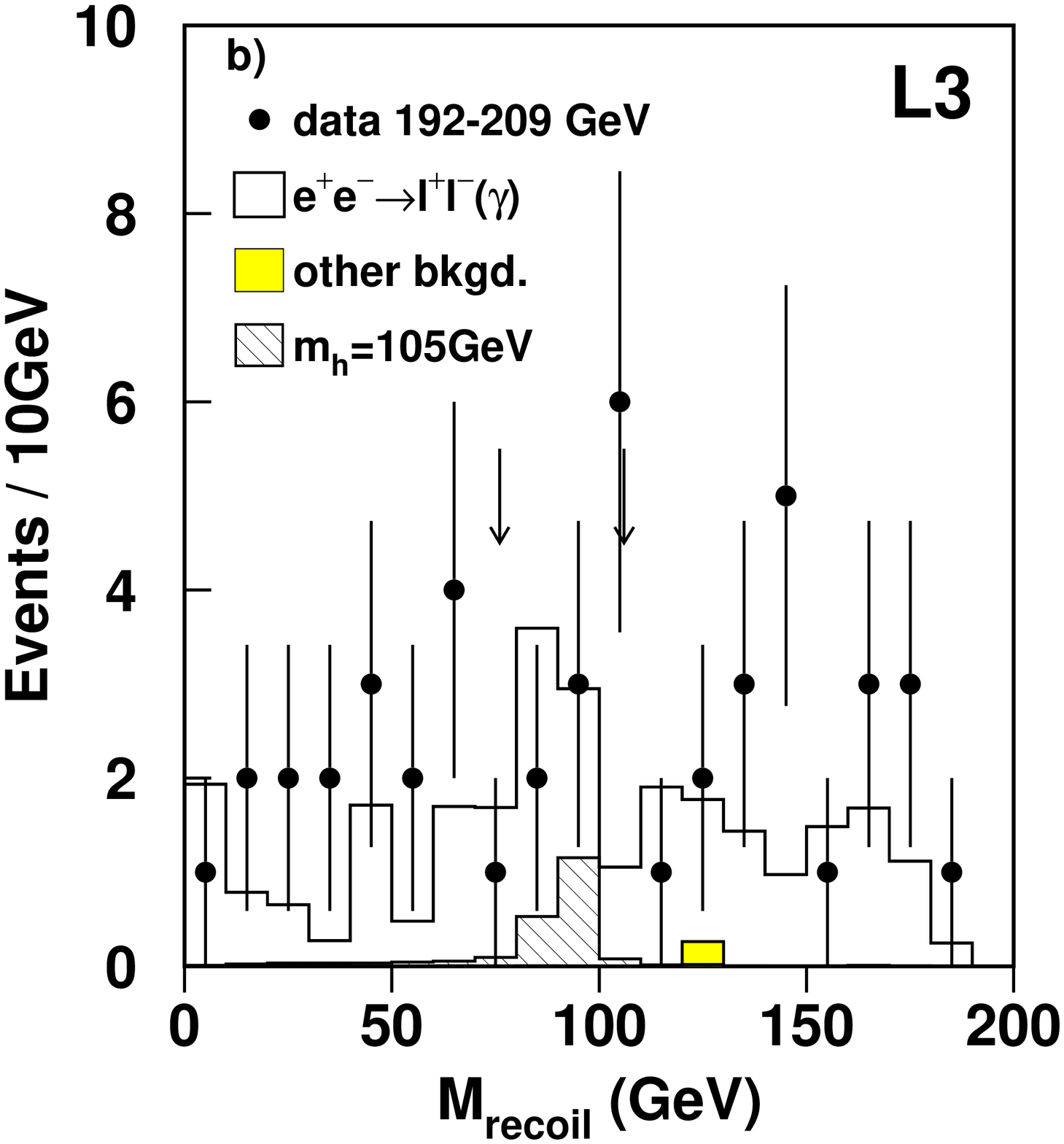} \\
\end{tabular} 
\vspace*{2cm}  
\caption{Distributions for the $\LL \gamma \gamma$ final state of
a) the energy of the second most energetic photon normalised
to the beam energy, for the preselected events, and
b) the recoil mass against the two photons.
Data, background and a
Higgs boson signal with mass $m_{\mbox{{\scriptsize h}}}$ =  105$\GeV$
and arbitrary normalisation are shown. 
The arrows indicate the values of the cuts.
}
\label{fig_lept}
\end{center}
\end{figure}

The distribution of the energy of the second most energetic photon 
normalised to the beam energy is presented 
in Figure \ref{fig_lept}a for the preselected events. 
Figure \ref{fig_lept}b shows the recoil mass against the photons 
after all selection criteria but the recoil mass.

The final selection requirement imposes the recoil mass 
to be consistent with the Z mass
within 15$\GeV$.
At centre-of-mass energies below $\sqrt s = 202\GeV$ the presence of two identified 
leptons is enforced. Their invariant mass is required to be between 81 and 101 GeV and 
the selection criterium on the recoil mass is relaxed.

The efficiency varies from 31$\%$ for a Higgs boson mass of 100$\GeV$ produced at $\sqrt s =192\GeV$, 
to 43$\%$ for a Higgs of 110$\GeV$ produced at $\sqrt s = 208\GeV$. The yield of this selection is presented in Table \ref{tab_sel2}.
7 events are observed in the data, with 8.0 expected from Monte Carlo, mainly from the 
$\mbox{e}^{+}\mbox{e}^{-} \to \LL (\gamma)$ processes.

\section*{Results}

\begin{figure}[H]
\begin{center}
\begin{tabular}{l}
\includegraphics[width=9.5cm]{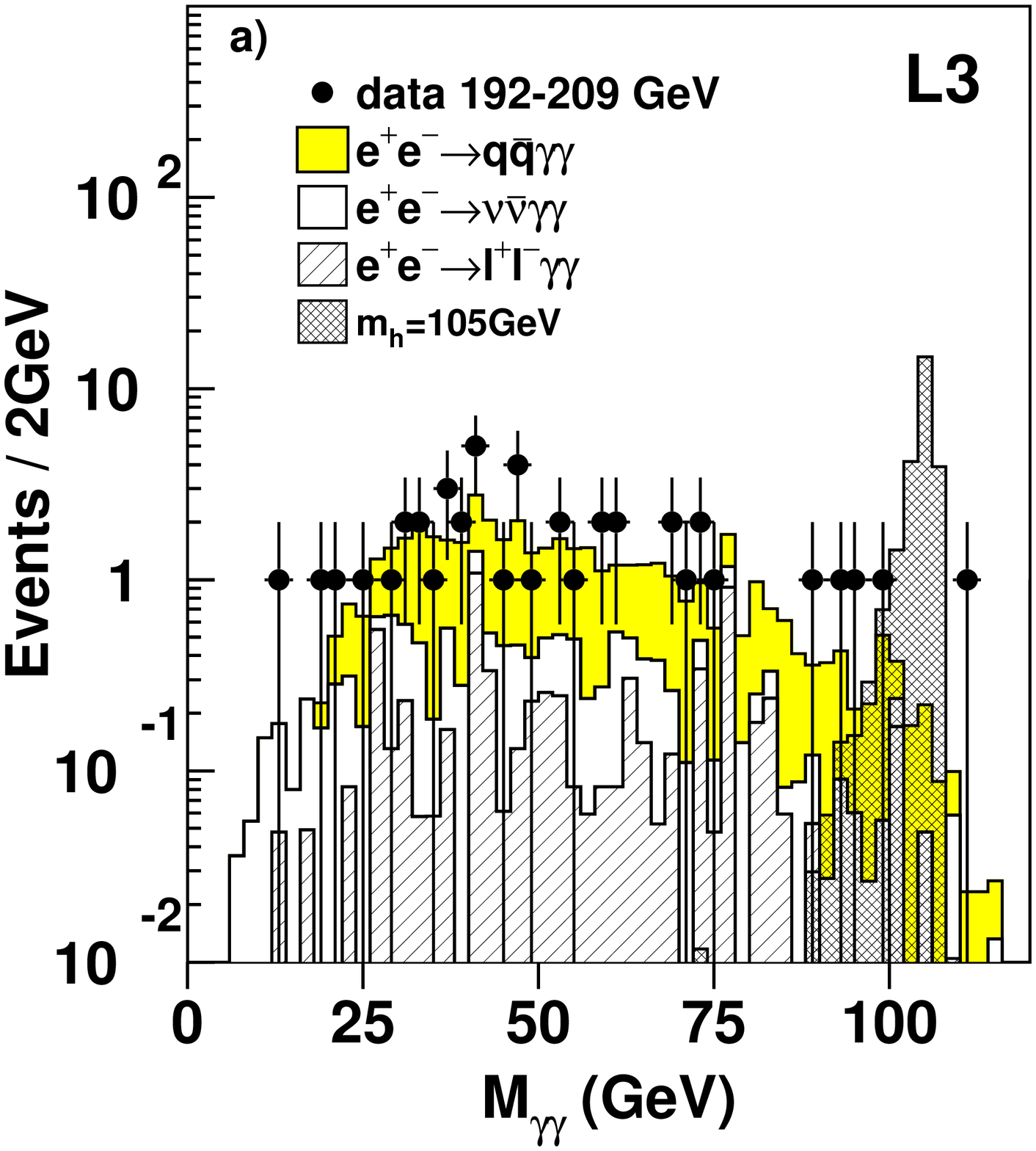} \\
\vspace*{-2cm}
\includegraphics[width=9.5cm]{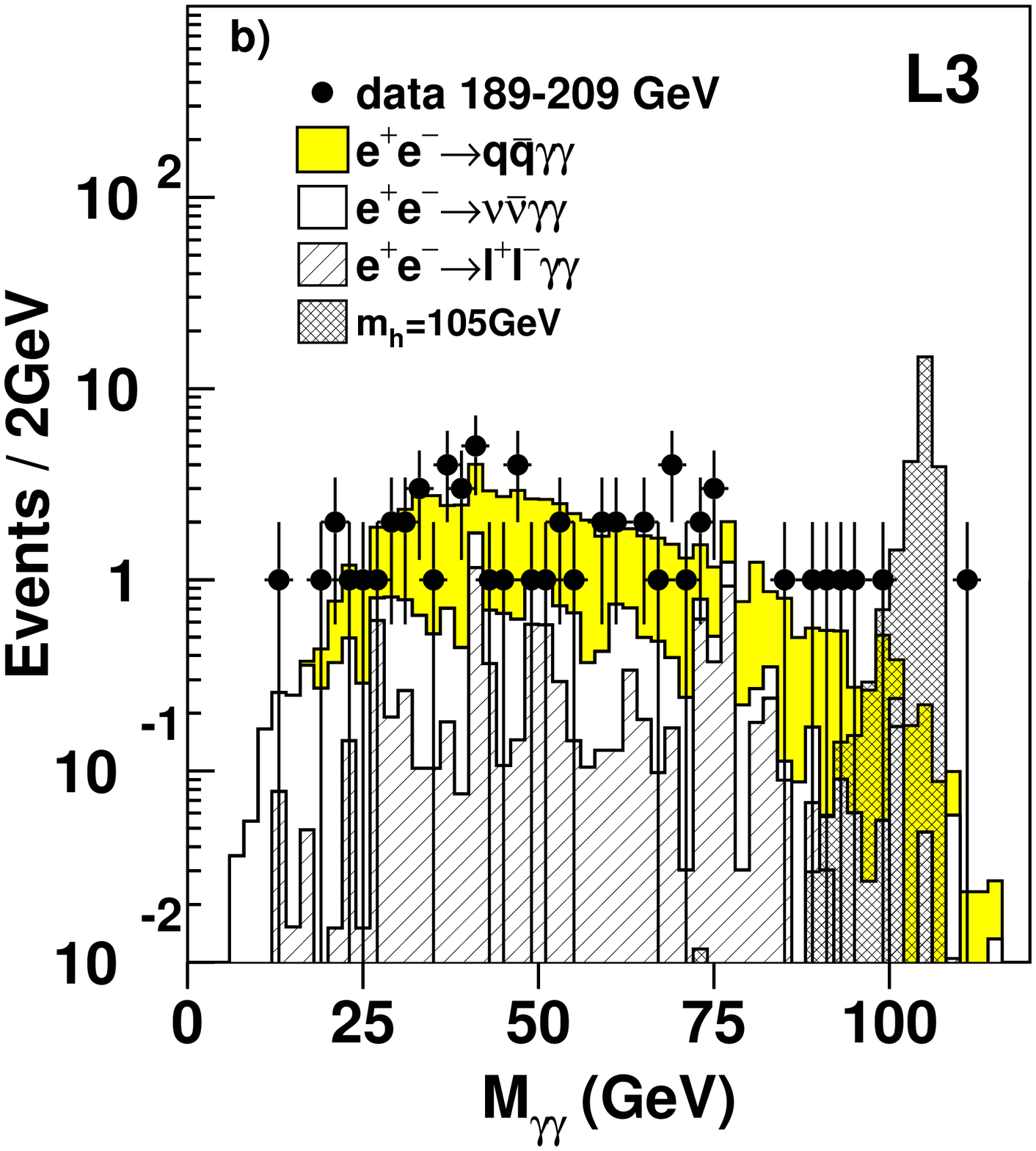} \\
\end{tabular}
\vspace*{2cm}
\caption{Distributions of the reconstructed di-photon invariant mass  
for all final states combined, after the
final selection. Data at a) $\sqrt s = 192-209\GeV$, and 
b) $\sqrt s = 189-209\GeV$ are shown together with the background and a
Higgs boson signal with mass $m_{\mbox{{\scriptsize h}}}$ =  105$\GeV$. The Standard Model cross section and a 
BR(h$\to \gamma \gamma$) = 1 are used.
}
\label{fig_res}
\end{center}
\end{figure}

No significant excess indicating the production of a Higgs boson decaying into two photons 
is observed in the data. 
The confidence level \cite{cl} for the absence of a Higgs signal 
is then calculated from the reconstructed di-photon invariant mass as the final discriminant variable.
This distribution is shown in Figure \ref{fig_res}a for the data analysed in this paper, 
collected at 
$\sqrt s = 192-209\GeV$ and in  Figure \ref{fig_res}b including the data 
collected at $\sqrt s = 189\GeV.$

The calculation of the limits takes into account systematic uncertainties of 2$\%$
on the signal expectations and 8$\%$ on the background.
The signal uncertainty follows from the Monte Carlo statistics that also accounts for a
4$\%$ uncertainty on the background.
Another uncertainty of 7\% is assigned to the background 
normalisation for hadronic events with photons. A variation of 2$\%$ 
of the calorimetric energy scale has little effect on the limits.    
The effects of the energy and angular resolutions of the 
photons and the systematic uncertainty on the integrated luminosity are also found to 
be negligible. 

\begin{figure}[H]
\begin{center}
\begin{tabular}{l}
\includegraphics[width=15cm]{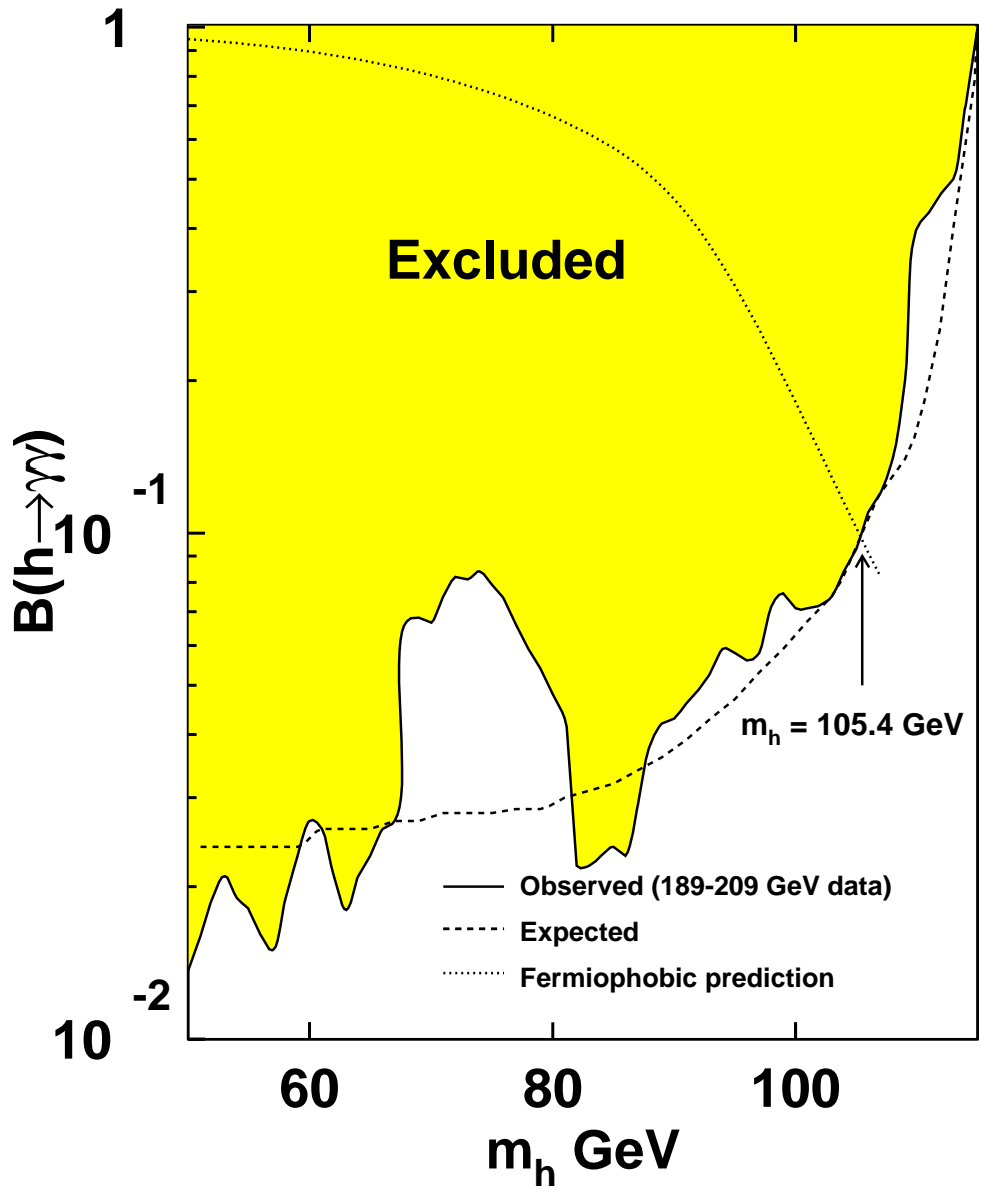} \\
\end{tabular}
\caption{Excluded values at 95$\%$ confidence level of $\mbox{BR}(\mbox{h}\to \gamma \gamma)$
as a function of the Higgs mass, in the assumption of the Standard Model production cross section.
The  expected 95$\%$ confidence level limit and the theoretical
prediction are also presented.
}
\label{fig_cl}
\end{center}
\end{figure}

Figure \ref{fig_cl} presents the upper limit on the branching fraction
$\mbox{BR}(\mbox{h} 
\to \gamma \gamma)$ as a function
of the Higgs mass, assuming the Standard Model cross section for the Zh production.
The
expected limit is also shown together with the
theoretical prediction for a fermiophobic Higgs boson as
calculated with the HDECAY program \cite{hdecay}.
Previous L3 results \cite{l3}
are included in the calculation of this limit. The observed limit for
$\mbox{BR}(\mbox{h}
\to \gamma \gamma)$~=~1 is 114$\GeV$. 
 
The lower limit on the mass of a fermiophobic Higgs boson is set at

\begin{center}
$m_{\mbox{{\scriptsize h}}} > 105.4$$\GeV$ at 95$\%$ confidence level,
\end{center}

\noindent
to be compared with the expected mass limit of 105.3$\GeV$.

\newpage
\typeout{   }     
\typeout{Using author list for paper 251 -- ? }
\typeout{$Modified: Jul 15 2001 by smele $}
\typeout{!!!!  This should only be used with document option a4p!!!!}
\typeout{   }
%
%
%
%
%
%

\newcount\tutecount  \tutecount=0
\def\tutenum#1{\global\advance\tutecount by 1 \xdef#1{\the\tutecount}}
\def\tute#1{$^{#1}$}
\tutenum\aachen            
\tutenum\nikhef            
\tutenum\mich              
\tutenum\lapp              
\tutenum\basel             
\tutenum\lsu               
\tutenum\beijing           
\tutenum\berlin            
\tutenum\bologna           
\tutenum\tata              
\tutenum\ne                
\tutenum\bucharest         
\tutenum\budapest          
\tutenum\mit               
\tutenum\panjab            
\tutenum\debrecen          
\tutenum\florence          
\tutenum\cern              
\tutenum\wl                
\tutenum\geneva            
\tutenum\hefei             
\tutenum\lausanne          
\tutenum\lyon              
\tutenum\madrid            
\tutenum\florida           
\tutenum\milan             
\tutenum\moscow            
\tutenum\naples            
\tutenum\cyprus            
\tutenum\nymegen           
\tutenum\caltech           
\tutenum\perugia           
\tutenum\peters            
\tutenum\cmu               
\tutenum\potenza           
\tutenum\prince            
\tutenum\riverside         
\tutenum\rome              
\tutenum\salerno           
\tutenum\ucsd              
\tutenum\sofia             
\tutenum\korea             
\tutenum\purdue            
\tutenum\psinst            
\tutenum\zeuthen           
\tutenum\eth               
\tutenum\hamburg           
\tutenum\taiwan            
\tutenum\tsinghua          

{
\parskip=0pt
\noindent
{\bf The L3 Collaboration:}
\ifx\selectfont\undefined
 \baselineskip=10.8pt
 \baselineskip\baselinestretch\baselineskip
 \normalbaselineskip\baselineskip
 \ixpt
\else
 \fontsize{9}{10.8pt}\selectfont
\fi
\medskip
\tolerance=10000
\hbadness=5000
\raggedright
\hsize=162truemm\hoffset=0mm
\def\r{\rlap,}
\noindent

P.Achard\r\tute\geneva\ 
O.Adriani\r\tute{\florence}\ 
M.Aguilar-Benitez\r\tute\madrid\ 
J.Alcaraz\r\tute{\madrid,\cern}\ 
G.Alemanni\r\tute\lausanne\
J.Allaby\r\tute\cern\
A.Aloisio\r\tute\naples\ 
M.G.Alviggi\r\tute\naples\
H.Anderhub\r\tute\eth\ 
V.P.Andreev\r\tute{\lsu,\peters}\
% ...
\newpage
\begin{list}{A}{\itemsep=0pt plus 0pt minus 0pt\parsep=0pt plus 0pt minus 0pt
                \topsep=0pt plus 0pt minus 0pt}
\item[\aachen]
 I. Physikalisches Institut, RWTH, D-52056 Aachen, FRG$^{\S}$\\
 III. Physikalisches Institut, RWTH, D-52056 Aachen, FRG$^{\S}$
\item[\nikhef] National Institute for High Energy Physics, NIKHEF, 
     and University of Amsterdam, NL-1009 DB Amsterdam, The Netherlands
\item[\mich] University of Michigan, Ann Arbor, MI 48109, USA
\item[\lapp] Laboratoire d'Annecy-le-Vieux de Physique des Particules, 
     LAPP,IN2P3-CNRS, BP 110, F-74941 Annecy-le-Vieux CEDEX, France
\item[\basel] Institute of Physics, University of Basel, CH-4056 Basel,
     Switzerland
\item[\lsu] Louisiana State University, Baton Rouge, LA 70803, USA
\item[\beijing] Institute of High Energy Physics, IHEP, 
  100039 Beijing, China$^{\triangle}$ 
\item[\berlin] Humboldt University, D-10099 Berlin, FRG$^{\S}$
\item[\bologna] University of Bologna and INFN-Sezione di Bologna, 
     I-40126 Bologna, Italy
\item[\tata] Tata Institute of Fundamental Research, Mumbai (Bombay) 400 005, India
\item[\ne] Northeastern University, Boston, MA 02115, USA
\item[\bucharest] Institute of Atomic Physics and University of Bucharest,
     R-76900 Bucharest, Romania
\item[\budapest] Central Research Institute for Physics of the 
     Hungarian Academy of Sciences, H-1525 Budapest 114, Hungary$^{\ddag}$
\item[\mit] Massachusetts Institute of Technology, Cambridge, MA 02139, USA
\item[\panjab] Panjab University, Chandigarh 160 014, India.
\item[\debrecen] KLTE-ATOMKI, H-4010 Debrecen, Hungary$^\P$
\item[\florence] INFN Sezione di Firenze and University of Florence, 
     I-50125 Florence, Italy
\item[\cern] European Laboratory for Particle Physics, CERN, 
     CH-1211 Geneva 23, Switzerland
\item[\wl] World Laboratory, FBLJA  Project, CH-1211 Geneva 23, Switzerland
\item[\geneva] University of Geneva, CH-1211 Geneva 4, Switzerland
\item[\hefei] Chinese University of Science and Technology, USTC,
      Hefei, Anhui 230 029, China$^{\triangle}$
\item[\lausanne] University of Lausanne, CH-1015 Lausanne, Switzerland
\item[\lyon] Institut de Physique Nucl\'eaire de Lyon, 
     IN2P3-CNRS,Universit\'e Claude Bernard, 
     F-69622 Villeurbanne, France
\item[\madrid] Centro de Investigaciones Energ{\'e}ticas, 
     Medioambientales y Tecnol\'ogicas, CIEMAT, E-28040 Madrid,
     Spain${\flat}$ 
\item[\florida] Florida Institute of Technology, Melbourne, FL 32901, USA
\item[\milan] INFN-Sezione di Milano, I-20133 Milan, Italy
\item[\moscow] Institute of Theoretical and Experimental Physics, ITEP, 
     Moscow, Russia
\item[\naples] INFN-Sezione di Napoli and University of Naples, 
     I-80125 Naples, Italy
\item[\cyprus] Department of Physics, University of Cyprus,
     Nicosia, Cyprus
\item[\nymegen] University of Nijmegen and NIKHEF, 
     NL-6525 ED Nijmegen, The Netherlands
\item[\caltech] California Institute of Technology, Pasadena, CA 91125, USA
\item[\perugia] INFN-Sezione di Perugia and Universit\`a Degli 
     Studi di Perugia, I-06100 Perugia, Italy   
\item[\peters] Nuclear Physics Institute, St. Petersburg, Russia
\item[\cmu] Carnegie Mellon University, Pittsburgh, PA 15213, USA
\item[\potenza] INFN-Sezione di Napoli and University of Potenza, 
     I-85100 Potenza, Italy
\item[\prince] Princeton University, Princeton, NJ 08544, USA
\item[\riverside] University of Californa, Riverside, CA 92521, USA
\item[\rome] INFN-Sezione di Roma and University of Rome, ``La Sapienza",
     I-00185 Rome, Italy
\item[\salerno] University and INFN, Salerno, I-84100 Salerno, Italy
\item[\ucsd] University of California, San Diego, CA 92093, USA
\item[\sofia] Bulgarian Academy of Sciences, Central Lab.~of 
     Mechatronics and Instrumentation, BU-1113 Sofia, Bulgaria
\item[\korea]  The Center for High Energy Physics, 
     Kyungpook National University, 702-701 Taegu, Republic of Korea
\item[\purdue] Purdue University, West Lafayette, IN 47907, USA
\item[\psinst] Paul Scherrer Institut, PSI, CH-5232 Villigen, Switzerland
\item[\zeuthen] DESY, D-15738 Zeuthen, 
     FRG
\item[\eth] Eidgen\"ossische Technische Hochschule, ETH Z\"urich,
     CH-8093 Z\"urich, Switzerland
\item[\hamburg] University of Hamburg, D-22761 Hamburg, FRG
\item[\taiwan] National Central University, Chung-Li, Taiwan, China
\item[\tsinghua] Department of Physics, National Tsing Hua University,
      Taiwan, China
\item[\S]  Supported by the German Bundesministerium 
        f\"ur Bildung, Wissenschaft, Forschung und Technologie
\item[\ddag] Supported by the Hungarian OTKA fund under contract
numbers T019181, F023259 and T024011.
\item[\P] Also supported by the Hungarian OTKA fund under contract
  number T026178.
\item[$\flat$] Supported also by the Comisi\'on Interministerial de Ciencia y 
        Tecnolog{\'\i}a.
\item[$\sharp$] Also supported by CONICET and Universidad Nacional de La Plata,
        CC 67, 1900 La Plata, Argentina.
\item[$\triangle$] Supported by the National Natural Science
  Foundation of China.
\end{list}
}
\vfill


\newpage

\newpage

\begin{table} [ht]
\begin{center}
\hspace*{-1.cm}
\begin{tabular}{|c|c c c c c c c c c c|}
\hline
$\sqrt s$ (GeV) &191.6 & 195.5 & 199.5 & 201.7 & 203.8& 205.1& 206.3& 206.6& 208.0& 208.6 \\
\hline
Luminosity (pb$^{-1}$) & 29.4&83.7&82.8& 37.0&7.6&68.1&66.9&63.7&8.2&0.1 \\
\hline
\end{tabular}
\caption{
The average centre-of-mass energies and corresponding integrated luminosities.
}
\label{tab_lumi}
\end{center}
\end{table}

\begin{table} [ht]
\begin{center}
\hspace*{-1.cm}
\begin{tabular}{|c|c|c||c|c|c|c|}\hline
                  &$N_{\rm D}$ &$N_{\rm B}$ &$\mbox{q}\bar{\mbox{q}}(\gamma)$& WW& 
$\mbox{Z}\mbox{e}^{+}\mbox{e}^{-}$& 
ZZ\\
\hline
Preselection
                              &17719&17739.2&11364.2&5876.7&139.4&358.9 \\
\hline
Selection
 &\phantom{0}\phantom{0}28&\phantom{0}\phantom{0}31.0&\phantom{0}
\phantom{0}30.5&\phantom{0}\phantom{0}\phantom{0}0.2&\phantom{0}0.1&
\phantom{0}\phantom{0}0.2    \\ 
\hline
\end{tabular}
\caption{
Number of events, $N_{\rm D}$, observed in data by the $\mbox{q}\bar{\mbox{q}} \gamma \gamma$
selection, compared with the Standard Model expectations, $N_{\rm B}$.
The Monte Carlo breakdown in different processes is given. 
}
\label{tab_sel1}
\end{center}
\end{table}

\begin{table} [ht]
\begin{center}
\hspace*{-1.cm}
\begin{tabular}{|c|c|c||c|c|c|c|}\hline
                  &$N_{\rm D}$&$N_{\rm B}$& $\mbox{e}^{+} \mbox{e}^{-} (\gamma)$& $\mu^{+} \mu^{-}(\gamma)$& 
$\tau^{+} \tau^{-}(\gamma)$& 
4 fermion \\
\hline
Preselection
                              &738&751.3&541.7&46.2&50.4&113.0 \\
\hline
Selection         
  &\phantom{0}7&\phantom{0}8.0&\phantom{0}4.2&\phantom{0}1.8&2.0&0.0   \\ 
\hline
\end{tabular}
\caption{
Number of events, $N_{\rm D}$, observed in data by the $\LL \gamma \gamma$
selection, compared with the Standard Model expectations, $N_{\rm B}$.
The Monte Carlo breakdown in different processes is given. 
}
\label{tab_sel2}
\end{center}
\end{table}

%

\newpage

%
\newpage

\newpage

\newpage

%
\newpage

\newpage

\newpage
\newpage
%
\vfill\newpage

\end{document}